\documentclass[conference,compsoc]{IEEEtran}
\ifCLASSOPTIONcompsoc
  \usepackage[nocompress]{cite}
\else
  \usepackage{cite}
\fi
\ifCLASSINFOpdf
\else
\fi
\newcommand{\attr}{\mathsf{times}}
\newcommand{\dice}{\mathsf{dice}}

\usepackage{amsmath,amssymb,amsthm,verbatim,hyperref}
\begin{document}
%
\title{Improving random walk rankings with feature selection and imputation \\ Science4cast competition, team Hash Brown}

\author{\IEEEauthorblockN{Ngoc Mai Tran}
\IEEEauthorblockA{Department of Mathematics\\ University of Texas at Austin\\ Austin, TX 78712\\
Email: tran.mai.ngoc@utexas.edu}
\and
\IEEEauthorblockN{Yangxinyu Xie}
\IEEEauthorblockA{Department of Computer Science\\
University of Texas at Austin\\ Austin, TX 78712\\
Email: yx4247@utexas.edu}}


%


\maketitle

\begin{abstract}
The Science4cast Competition consists of predicting new links in a semantic network, with each node representing a concept and each edge representing a link proposed by a paper relating two concepts. This network contains information from 1994-2017, with a discretization of days (which represents the publication date of the underlying papers).
Team Hash Brown's final submission, \emph{ee5a}, achieved a score of 0.92738 on the test set. Our team's score ranks \emph{second place}, 0.01 below the winner's score. This paper details our model, its intuition, and the performance of its variations in the test set.
\end{abstract}


%
\IEEEpeerreviewmaketitle

\section{Background}

Predicting future trends in science by completing semantic networks using machine learning approaches has drawn a lot of attention in recent years \cite{s4s, wang_barabasi_2021, evans2011metaknowledge, rzhetsky2015choosing, krenn2020predicting}. Last year, Krenn and Zeilinger introduced SEMNET, a deep neural network model using 17 features extracted from the semantic network in the context of quantum physics \cite{krenn2020predicting}. These features include graph-theoretic heuristics such as common neighbors, degree centrality, and distance, as well as features based on the nature of the task, such as the number of papers that a concept occurs. 

A similar kind of semantic network is the collaboration network. Recently, Wang et. al. published a benchmark dataset, \texttt{ogbl-collab}, which contains an undirected graph, representing a subset of the collaboration network between authors indexed by Microsoft academic graph \cite{wang2020microsoft, hu2020open}. In this network, each node represents an author, and edges indicate the collaboration between authors. 

Multiple competing approaches have been proposed and tested on this benchmark dataset. On the one hand, the classical graph-theoretic heuristics such as the Adamic Adar Index \cite{adamic2003friends} and common neighbors alone proved to be powerful predictors. On the other hand, novel graph convolutional network approaches, such as graphSAGE \cite{hamilton2017inductive, singh2021edge} and SEAL (learning from Subgraphs, Embeddings, and Attributes for Link prediction) \cite{zhang2018link, zhang2020revisiting} also gained competitive performance. More recently, one recommender-systems-inspired approach, the HOP-Rec node embedding method, has achieved the state-of-the-art performance on this benchmark \cite{yang2018hop}.

\section{The model}

\subsection{Transformation}
The raw data consists of a list of tuples of the form $(u,v,d)$, which means nodes $u$ and $v$ were investigated together in some paper that appeared on day $d$. We take the quantile transform of days to transform the data into the form $(u,v,t)$ for $t \in (0,1)$. Let $\mathcal{G}_{[t_a,t_b]} = (\mathcal{V}_{[t_a,t_b]},\mathcal{E}_{[t_a,t_b]})$ denote the network formed by edges that appeared within the time interval $[t_a,t_b]$. We also write $\mathcal{G}_t$ for $\mathcal{G}_{[0,t]}$, and similarly for $\mathcal{V}$ and $\mathcal{E}$. Each edge $(u,v) \in \mathcal{G}_t$ has edge attribute $$\attr_t(u,v) = \{s \in (0,1): s < t, (u,v,s) \in \mbox{ dataset}\}.$$ 
This is the list of time stamps that the edge $(u,v)$ appeared in the dataset before time $t$. 
Similarly, the nodes have attributes 
\begin{align*}
&\attr_t(u) = \\ 
&\{s \in (0,1): s < t, (u,v,s) \in \mbox{ dataset for some node } v\},    
\end{align*}
which is the list of times that the node $u$ appeared in the dataset before time $t$. From the original undirected graph $\mathcal{G}_t$, we also define a directed graph $\vec{G}_t$, where $u \rightarrow v$ means node $u$ appeared in the dataset before node $v$, that is, $\min(\attr_t(u)) \leq \min(\attr_t(v))$. 

\subsection{Model overview}

We use a feature embedding model. The premise is that the probability that nodes $u$ and $v$ form an edge in the future, is a function of the node features $f(u)$, $f(v)$, and some edge feature $h(u,v)$. That is,
$$\mathbb{P}(u \sim v) = \mathcal{F}(f(u),f(v), h(u,v)).$$
We hand-engineered the feature functions $f$ and $h$ (cf. Section \ref{sec:features.definitions}). Based on past work on link predictions on citation networks \cite{krenn2020predicting}, we chose node features $f$ that capture popularity at the current time $t_0$ (such as degree and PageRank \cite{brin1998anatomy}). We also use these features' first and second time-derivatives to capture the evolution of the node's popularity over time. Meanwhile, $h$ is a measure of similarity between nodes. For the classifier $\mathcal{F}$, we implemented logistic regression, multilayer perceptron (MLP) with ReLU activation. We set the logistic regression as the benchmark and used the learned coefficients for feature selection. The final submission has 31 node features for each node, and two edge features: the HOP-rec score and the Dice similarity score. This gives $31 \times 2 + 2 = 64$ features in total. All of the 62 node features are standardized to have mean 0 and variance 1. The best-performing classifier on the test set is learned by an MLP with architecture $(13,13,13,13,13)$. We note that our model's performance is quite robust to variations in the features as well as the MLP architecture (cf. Table \ref{tab:overall comparison}). In particular, logistic regression with the same feature set achieves a test score of over 0.925, which would also qualify for second place.

\subsection{Creating the training set and dealing with cold start}

For training, we set $t_0 = 0.9$. With $\Delta t = 0.15$, each node $u \in \mathcal{V}_{t_0}$ gets a feature vector $f(u) \in \mathbb{R}^{31}$. For positive training examples, we took all edges $(u,v)$ between nodes $u,v \in \mathcal{V}_{t_0}$ that appeared for the first time after time $0.9$. For negative training examples, we sampled an equal number of random pairs $(u,v)$ for $u,v \in \mathcal{V}_{t_0}$ such that the edge $(u,v)$ is not in the final graph (at time $1$). We added extra edges to mimic the imputation procedure for the cold start (see below). This made up the training data. Then we applied stratified-$k$-fold to create training and validation sets. In the final model, we fitted the MLP on 75\% of the available training data. We say a node $v$ is {\bf seen} if it appeared in the training data, i.e. $v \in \mathcal{V}_{t_0}$, and {\bf unseen} otherwise.

Cold start is the problem that some nodes in the test set do not appear in the training set. This problem is quite acute in the 2017 test set: about 3\% of the test pairs consist of unseen nodes, and another 33\% of the test pairs consist of an unseen node and a seen node. Our strategy for cold start is imputation. The idea is that an unseen node is simply a node born in the future, so its features should look like a recently born node in the training set. Therefore, we define the `newborn' nodes $Newborn(t_0)$ to be the set of nodes $u \in \mathcal{V}_{t_0}$ that appeared for the first time after time $t_0 - 0.1$; i.e. $u \in (\mathcal{V}_{t_0}\setminus \mathcal{V}_{t_0-0.1})$. If $u$ is unseen, then $f(u)$ is the average of the features of the newborns. That is, we define
$$f(u) := \mathbb{E}_{u' \in \mathcal{V}_{t_0}}f(u').$$ 
If both $u,v$ are unseen, then the edge feature is also averaged over pairs from this set, that is, 
\begin{equation}\label{eqn:h.impute.blind}
h(u,v) := \mathbb{E}_{u',v' \in \mathcal{V}_{t_0}, u' \neq v'} h(u',v').  
\end{equation}
If $u$ is unseen but $v$ is seen, then 
\begin{equation}\label{eqn:h.impute.pirate}
h(u,v) := \mathbb{E}_{u'\in \mathcal{V}_{t_0}} h(u',v).     
\end{equation}
In the training set, with $t_0 = 0.9$, we added a small fraction (7\%) of edges between unseen-seen pairs. We found that with imputation during training, the test AUC scores across all models consistently increased by about 0.02. A major reason why we settled for HOP-rec and Dice similarity scores for $h$ is that \eqref{eqn:h.impute.blind} and \eqref{eqn:h.impute.pirate} can be computed quickly for the imputation step, without having to iterate over all possible pairs between newborn nodes and the seen nodes. 

\subsection{Features Definitions}\label{sec:features.definitions}

{\bf Full feature set}
With respect to a fixed time $t_0$ and a time derivative parameter $\Delta t$, each node $u \in \mathcal{G}_{t_0}$ has the following features. 

1. $\min(\attr_{t_0}(u))$: the first time that node $u$ appears. 

2. The number of unique time stamps of $\attr_{t_0}(u)$, log-transformed. This is a rough estimate for the number of papers that node $u$ appeared in up to time $t_0$. 

3. Clustering coefficient \cite{holland1971transitivity, watts1998collective}, degree (log-transformed), and PageRank (log-transformed) \cite{brin1998anatomy} of $u$ in $\mathcal{G}_{t_0}$

4. The log-transformed in-degree and out-degree of $u$ in $\vec{\mathcal{G}}_{t_0}$

5. The first and second discrete-time derivatives of the features listed in 2. to 4. in forward-time and backward-time. For a feature $f(u,\mathcal{G}_{t_0}) \in \mathbb{R}$ (for example, page rank), the first forward-time derivative is defined as
$$ f(u,\mathcal{G}_{[0,t_0]}) - f(u,\mathcal{G}_{[0,t_0 - \Delta t]}), $$
while the first backward-time derivative is defined as
$$ f(u,\mathcal{G}_{[0, t_0]}) - f(u,\mathcal{G}_{[\Delta t, t_0]}). $$
Similarly, the second forward-time derivative is
$$ f(u,\mathcal{G}_{[0,t_0]}) - 2f(u,\mathcal{G}_{[0,t_0 - \Delta t]}) + f(u,\mathcal{G}_{[0,t_0-2\Delta t}]), $$
while the second backward-time derivative is 
$$ f(u,\mathcal{G}_{[0,t_0]}) - 2f(u,\mathcal{G}_{[\Delta t,t_0]}) + f(u,\mathcal{G}_{[2\Delta t, t_0]}). $$

Altogether, this gives $1 + 4 \times 5 + 2 \times 5 = 31$ features for each node at a fixed time $t_0$. 

Let $f_{t_0}(u) \in \mathbb{R}^{31}$ denote the feature vector for node $u$. The feature vector of a node pair $(u,v)$ at time $t_0$ is obtained by concatenating the element-wise min and max of $f_{t_0}(u)$ and $f_{t_0}(v)$ (which gives $31 \times 2 = 62$ features), and then we append the edge features, the HOP-rec score and Dice similarity score, of the pair $(u,v)$. Both HOP-rec and Dice scores are computed with respect to the graph at time $t_0$. Altogether, this gives $64$ features for each pair $(u,v)$, where $u,v \in V_{t_0}$, that is, both $u$ and $v$ are seen nodes by time $t_0$. 


{\bf Dice Similarity}
For a node $u$ in a graph $G$, let $N(u)$ denote its neighborhood, that is, the set of nodes incident to $u$. The dice similarity of a pair of nodes $(u,v)$ is
$$ \dice(u,v) = \frac{2|N(u) \cap N(v)|}{|N(u)| + |N(v)|}. $$
This number has a probabilistic meaning as follows. Pick a node at random from the set $N(u)$ or $N(v)$. Then $\dice(u,v)$ is the probability that this node is incident to both $u$ and $v$. Thus, dice similarity can be generalized between sets $U, V$ as follows: pick a node $u$ at random in $U$ and a node $v$ at random in $V$, then pick a node $w$ at random from the set $N(u)$ or $N(v)$. Then, $\dice(U,V)$ is the probability that $w$ is incident to both $u$ and $v$. 
Our definition is \emph{different} from the original Sorensen-Dice coefficient defined for pair of sets \cite{sorensen1948method}. However, it is  consistent with our imputation methods for edge features in \eqref{eqn:h.impute.blind} and \eqref{eqn:h.impute.pirate}. Moreover, it is easy to compute by block matrix operations on the adjacency matrix, since 
\begin{equation}\label{eqn:dice.pirate}
d(u,V) = \frac{2\sum_{v \in V}|N(u) \cap N(v)|}{|N(u)| \times |V| + \sum_v |N(v)|}, 
\end{equation}
and
\begin{equation}\label{eqn:dice.blind}
d(V,V) = \frac{2\sum_{w \in V: |N(w) \cap V| \geq 2} \binom{|N(w) \cap V|}{2}}{\sum_{w \in V}|N(w) \cap V|\times(|V|-1)}.    
\end{equation}

{\bf HOP-rec score} HOP-rec, recently introduced in \cite{yang2018hop}, is a ranking algorithm based on random walks. For each node in the training graph, the HOP-rec algorithm outputs a high-dimensional embedding vector, such that nodes within a short distance of each other tend to have higher cosine similarity scores. For each pair of nodes, the HOP-rec score is the cosine similarity of their corresponding HOP-rec embeddings. On the \texttt{ogbl-collab} dataset, \cite{Lin2021} reported better results when older papers are removed from the dataset. Following this idea, we removed all edges that appeared \emph{before} time $t = 0.5$ (day 7891, corresponds to August 11th, 2011), and computed the HOP-rec embedding for the remaining graph. This removed 6.32\% of unique edges (not counting multiplicity) and 1.27 \% of nodes. These are edges and nodes that have not re-appeared between time $t =0.5$ and $t = 1$. We imputed HOP-rec scores for edges involving unseen nodes using \eqref{eqn:h.impute.blind} and  \eqref{eqn:h.impute.pirate}, where the expectation is computed as the empirical average.  

{\bf Implementations} 
For PageRank and Dice similarity for known nodes, we used the implementation of  provided by the Python package \texttt{igraph} \cite{igraph}. For Dice similarity imputation, we used the formulas in \eqref{eqn:dice.pirate} and \eqref{eqn:dice.blind}. For HOP-rec embeddings, we used the code provided by \cite{smore} and \cite{Lin2021}. 

{\bf Reproducibility} 
Since the output of HOP-rec is random, we saved the HOP-rec embeddings. Re-runs of the HOP-rec embeddings produced an insignificant change in cosine similarity scores between the pairs (paired t-test on 1000 random pairs give $t = 1.1712323$, $p$-value = 0.241). For the classifiers (logistics and MLPs), we used the implementations from the Python package \texttt{scikit-learn} \cite{sklearn} with fixed seeds. So, given the HOP-rec embeddings, the final model parameters and results are fully reproducible. Reruns of our code produce the same ranking vector for $99.999\%$ of the 1 million entries. The discrepancies between runs are less than 100 edges, and these edges always appeared at the tail (between 900,000 and 1 million in ranking). The predicted probabilities for these edges hit close to machine precision, so this is simply a numerical precision error. 

The codes to reproduce the final model, along with the HOP-rec embeddings and model parameters, are available at \textbf{\texttt{github.com/princengoc/s4s-final}}. We used a CPU with Ryzen 5700G, 8 cores, 64GB of RAM. From start to finish, the final model took approximately 10 minutes to reproduce. The bulk of this time is taken by data-processing and feature generations. Re-runs of the model, with the features all produced, took less than 5 minutes per run. 

\subsection{Alternative methods and features that we explored}

We also experimented with the following methods. The best results for each model are reported in \eqref{tab:overall comparison}. As seen, some of these methods do improve upon our vanilla model. However, the improvements are minor and strictly below that of our final model. 

{\bf Graph-Neural-Network-based embeddings}
\begin{enumerate}
    \item {\bf Graph Convolutional Networks}(GCN) \cite{kipf2016semi} borrows the concept of convolution from the convolutional neural network (CNN) and convolve the graph directly according to the connectivity structure of the graph as the filter to perform neighborhood mixing. 
    \item {\bf GraphSAGE} \cite{hamilton2017inductive}, unlike GCN, is an inductive learning algorithm that learns aggregator functions, which can induce the embedding of a node given its features and neighborhood information.
\end{enumerate}
For a complete comparison between these two graph neural networks, see \cite{hu2020open}. Our implementation of both embeddings are adapted from the Open Graph benchmark \cite{hu2020open} and the \texttt{PyTorch Geometric} library \cite{fey2019fast}. 

{\bf Weighted Approximate-Rank Pairwise}(WARP) \cite{weston2011wsabie, weston2013learning}, originally developed for recommender systems, works by sampling $N$ positive items, ordering them by the score assigned by the model, and then weighting the example as a function of this ordered set. We adapted the code provided by \cite{smore} for this embedding.

{\bf Dropout}: During training, one randomly zeroes some of the elements of the input tensor with probability $p$, which we denote as the dropout rate, using samples from a Bernoulli distribution. Each channel will be zeroed out independently on every forward call. This is a common regularization technique \cite{hinton2012improving}.

\section{Experiment setup}
The provided {\bf training} dataset contains approximately 8,400 snapshots of the growing semantic network – one snapshot for each day from the beginning of 1994 to the end of 2017. The {\bf test} dataset contains a list of 1,000,000 vertex pairs that are unconnected by 2017. The positive samples in the test dataset contain a subset of the future links formed between 2017-2020 in the semantic network, which do not exist yet in 2017. The competition matrix is the  \textbf{AUC score}, which is the area under the ROC curve \cite{hanley1982meaning}. 

\section{Results and conclusion}\label{sec:results}

A direct comparison of the results of all competing approaches on the test set are summarized in Table \ref{tab:overall comparison}. 

\begin{table}[h]
\centering
\begin{tabular}{||c|c||} 
\hline
Models & Test\\ 
\hline\hline 
HOP-rec only & 0.89 \\
Logistic baseline & 0.9250 \\
MLP (32x3) & 0.92678 \\ 
MLP (13x5) & \textbf{0.92739} \\
MLP (13x5) + WARP & 0.92718 \\
MLP (13x5) + dropout & 0.9257 \\
MLP (13x5) + SAGE & 0.92605 \\ 
MLP (13x5) + GCN & 0.9252 \\ 
\hline 
\end{tabular}
\vskip12pt
\caption{Performances of different models on the test set.}
\label{tab:overall comparison}
\end{table}

Overall, the performance of our models is consistently between 0.9247 and 0.9274 and is significantly above the plain HOP-rec model. In particular, the simplest model is logistic regression, which would have qualified for second place as well. Among the features of the logistic regression, HOP-rec scores and PageRank have the largest weights, indicating that random-walk-based ranking models perform well for this link prediction task. 

In summary, we explored various feature embedding models for this link prediction problem and consistently got test AUC between 0.9247 and 0.9274. Our best model (0.9274) was an MLP with 5 hidden layers of width 13 each. Overall, we found that imputation and good features engineering are essential.

\ifCLASSOPTIONcompsoc
  \section*{Acknowledgments}
\else
  \section*{Acknowledgment}
\fi

The authors are supported by NSF Grant DMS-2113468, the NSF IFML 2019844 award to the University of Texas at Austin, and the Good Systems Research Initiative, part of University of Texas at Austin Bridging Barriers.



%

\bibliography{mybibfile}
\bibliographystyle{plain}

\end{document}